# Interaction of relativistically intense laser pulses with long-scale near critical plasmas for optimization of laser based sources of MeV electrons and gamma-rays

O N Rosmej [1, 2], N E Andreev [3, 4], S Zaehter [2], N Zahn [2], P Christ [2], B Borm [1,2], T Radon [1], A Sokolov [1], L P Pugachev [3, 4], D Khaghani [5], F Horst [1, 6], N G Borisenko [7], G Sklizkov [7], V G Pimenov [8]

[1] Helmholtzzentrum GSI-Darmstadt, Planckstr.1, 64291 Darmstadt, Germany
[2] Goethe University, Frankfurt, Max-von-Laue-Str. 1, 60438 Frankfurt am Main, Germany
[3] Joint Institute for high Temperatures, RAS, Izhorskaya st.13, Bldg. 2, 125412 Moscow, Russia
[4] Moscow Institute of Physics and Technology (State University), Institutskiy Pereulok 9, 141700 Dolgoprudny, Moscow Region, Russia
[5] Institute of Optics and Quantum Electronics, Abbe Center of Photonics, Friedrich Schiller University Jena, Max-Wien-Platz 1, 07745 Jena, Germany
[6] THM University of Applied Sciences, Wiesen Str. 14, 35390 Giessen, Germany
[7] P. N. Lebedev Physical Institute, RAS, Leninsky Prospekt 53, 119991 Moscow, Russia
[8] N.D. Zelinskiy Institute of Organic Chemistry RAS, Leninsky Prospekt 47, 119991 Moscow, Russia

E-mail: o.rosmej@gsi.de

**Abstract**

Experiments were performed to study electron acceleration by intense sub-picosecond laser pulses propagating in sub-mm long plasmas of near critical electron density (NCD). Low density foam layers of 300-500 μm thickness were used as targets. In foams, the NCD-plasma was produced by a mechanism of a super-sonic ionization when a well-defined separate ns-pulse was sent onto the foam-target forerunning the relativistic main pulse. The effect of the relativistic laser pulse channeling and creation of quasi-static azimuthal magnetic and radial electric fields that keeps electrons in the channel ensured effective coupling of the laser energy into energetic electrons. Application of sub-mm thick low density foam layers provided substantial increase of the electron acceleration path in a NCD-plasma compared to the case of freely expanding plasmas created in the interaction of the ns-laser pulse with solid foils. Performed experiments on the electron heating by a 100J, 750 fs short laser pulse of $2-5\times10^{19}$ W/cm$^2$ intensity demonstrated that the effective temperature of supra-thermal electrons increased from 1.5-2 MeV, in the case of the relativistic laser interaction with a metallic foil at high laser contrast, up to 13 MeV for the laser shots onto the pre-ionized foam. The observed tendency towards the strong increase of the mean electron energy and the number of ultra-relativistic laser-accelerated electrons is reinforced by the results of gamma-yield measurements that showed a 1000-fold increase of the measured doses. The experiment was supported by the 3D-PIC and FLUKA simulations made for used laser parameters and geometry of the experimental set-up. Both measurements and simulations show high directionality of the acceleration process, since the strongest increase in the electron energy, charge and corresponding gamma-yield was observed close to the direction of the laser pulse propagation. The charge of super-ponderomotive electrons with E > 30 MeV reaches a very high value of 78nC.

Keywords: relativistically intense laser pulses, NCD-plasmas, low density polymer aerogels, super-sonic ionization, super-ponderomotive electrons, gamma-rays.

## 1. Introduction

Experimental investigation of high energy density (HED) matter states created with intense laser [1, 2,] and heavy ion beams [3, 4] requires active backlighting with highly penetrating gamma-rays and energet-ic particles that provide important diagnostic tools to access plasma parameters and structural information from inside the high areal density samples. Intense and well directed beams of photons with energies far above 100 keV and tens to hundreds of MeV energetic particle beams of electrons/protons are the best candidates for such radiographic applications. TW and PW-class lasers systems that deliver laser pulses of relativistic intensities are widely used to generate relativistic particle beams and gamma-radiation. Micrometer-small size of the laser focus on a target surface together with a short laser pulse duration ensure high radiographic potential of theses secondary sources, providing up to 10 micrometer spatial resolution and snap shots of the WDM-object density distribution in ps- up to fs-time scale, short enough for analyses of the hydrodynamic motion of heated matter [5, 6].

Electrons play a major role at the very first stage of laser-matter interaction that lead to production of laser based sources of radiation and particles. There are different mechanisms of laser energy transfer to high energy electrons depending on the laser parameters and the type of targets from solid density targets with sharp boundary to extended low density gas targets. In solids, the mechanism strongly depends on gradients of the preplasma on the target surface and can be the vacuum/Brunel [7], resonant absorption in critical density, the ponderomotive or (JxB) mechanism of acceleration [8, 9], stochastic heating [10-13] etc.

Laser interaction with low density gas targets provides effective acceleration of electrons to high energies in the wakefields generated in plasma channels [14-16]. Great results were achieved in generation of monoenergetic electron beams with energies from hundreds of MeV up to several GeV in experiments on interaction of relativistic laser pulses with low density gas jets and capillary plasmas, see e.g. [17-20]. Nevertheless, the charge carried by these electron beams does not exceed tens pC, not sufficient to radiograph HED-samples in experiments with a high level of background radiation.

One of the possibilities to increase the electron beam charge above a nano-Coulomb (nC) level keeping the electron energy at a level of tens up to hundreds of MeV, is to use the advantage of relativistic laser interaction with plasmas of subcritical and near critical density (NCD) [21-24]

One of the first theoretical works that discusses particle acceleration in relativistic laser channels generated in near critical plasmas is based on results of 3D PIC-simulations [25]. Simulations demonstrated effects of channeling and filamentation of the relativistic laser pulse in the NCD-part of expanding plasma and generation of a strong current of energetic, 10-100 MeV, electrons that have Boltzmann-like energy distribution with an effective temperature, which depends among others on the laser pulse intensity and the length of the NCD plasma region. This strong electron current is accompanied by the creation of a giant azimuthal quasi-static magnetic field [28, 30]. The mechanism of the electron acceleration in NCD plasmas has intrinsically complex nature, as it involves many physical processes simultaneously. In [25] the authors propose a mechanism of the direct laser energy coupling into hot electrons that occurs in relativistic laser channels. This coupling requires strong self-generated static electric and magnetic fields that confine fast electrons in relativistic channels. Electrons experience transverse betatron oscillations that provide efficient energy exchange when the betatron frequency becomes close to the Doppler shifted laser frequency [25]. The effective electron temperatures obtained numerically for the case of the relativistic laser interaction with expanding plasmas, described by the exponential electron density profile with the scale length $L$=30 μm was 4.5 MeV for $I_L$=$10^{19}$ W/cm$^2$ and 14 MeV for $I_L$=$10^{20}$ W/cm$^2$.

After A. Pukhov [25] extended analyses of the relativistic laser pulse interaction with sub-critical plasmas was made by A. Arefiev and V. Khudik et al [26, 27]. They examine the processes of direct laser acceleration (DLA) of relativistic electrons undergoing betatron oscillations in a plasma channel and the role played by transverse and longitudinal quasi-static electric fields. In [27] a universal scaling for the maximum attainable electron energy was derived analytically; the authors showed as well the threshold dependence of the final energy gain on the laser intensity.

Up to now, only few experiments have been performed to demonstrate the advantages of the discussed mechanism of the electron acceleration. The energy transfer from an ultra-intense laser pulse with intensity $10^{20}$ W/cm$^2$ to hot electrons in NCD plasmas was investigated in dependence on the pre-plasma scale length [29] partially experimentally and partially using 2D PIC-simulations. In order to produce one-dimensional expansion of the plasma with a well-controlled scale length, a separate 5 ns long laser pulse with a 100 μm large focal spot was used. In the experiment, the coupling of the energy of the ultra-intense laser pulse into hot electrons was analyzed indirectly using measurements of Cu Kα-intensity and proton spectra in dependence on the pre-plasma scale length that in turn was simulated in 1D approximation for different energies of the long pulse. The energy distribution of energetic electrons was not measured directly but simulated using a 2D PIC-code. A discovered one order of magnitude of variation in the coupling efficiency of the laser energy into fast electrons was explained by the existence of a density gradient optimum that ensures strong laser pulse self-focusing and channeling processes that drive energy absorption over the extended length in performed plasmas.

Measurements of electrons accelerated by a relativistic laser pulse propagating across a mm-long extended sub-NCD plasma plume were reported in [22]. Experiments showed a strong increase of the "temperature" and number of supra-ponderomotive electrons caused by the increased length of a relativistic ion channel.

Production of a hydrodynamically stable NCD-plasma layer remains an important issue for such type of experiments. 3D PIC simulations of the relativistic laser interaction with large-scale NCD plasmas [30, 31] demonstrated that low density CHO aerogel [32-34] is a very prospective material for creation of sub-mm long NCD plasmas and efficient electron acceleration. Same conclusion can be made from the recently published experiments on the Omega EP-laser [24] where pre-ionized by a ns ASE-prepulse 250 μm thick foams with densities from 3 up to 100 mg/cm$^3$ ($n_e$ = 0.9-30×10$^{21}$cm$^{-3}$) were irradiated by a laser pulse of 1 kJ energy, 8-10 ps duration and 5.3±1.8×10$^{19}$ W/cm$^2$ intensity. Approximation of the high energy tail of the measured electron spectra with a Maxwell-like function resulted in 2 up to 4 times higher electron temperatures than $U_{pond} \simeq$ 5.4 MeV, corresponding to an $a_L$=6.5.

In this article we present new experimental results on the interaction of relativistic sub-picosecond laser pulses with sub-mm long NCD plasmas. Experiments were carried out with a 1.053 μm linearly polarized laser pulse of 80-100 J energy and 750±250 fs pulse duration. Resulting peak vacuum intensity reached 2.1-5.1×10$^{19}$ W/cm$^2$. The polymer foam layers of 2 mg/cm$^3$ density and up to 500 μm thickness were used to create hydrodynamically stable, large scale, quasi-homogeneous plasmas targets. The NCD-plasma was produced by a mechanism of a super-sonic ionization when a well-defined ns-pulse was sent onto the foam target forerunning the relativistic main pulse. Energy and duration of the ns-pulse were well matched to the mean volume density and the thickness of foam layers in order to optimize the velocity of the supersonic ionization wave that creates high aspect ratio NCD plasma. Direct measurements of the electron energy distribution were performed by means of two electron spectrometers with a static magnetic field. By comparison of shots onto metallic foils and onto pre-ionized low density foam layers it was demonstrated that an effective temperature of supra-thermal electrons increased from 1.5-2.2 MeV, in the case of the relativistic laser interaction with a metallic foil at high laser contrast, up to 13 MeV for the laser shots onto the pre-ionized 300 and 500 μm thick foam layers. Strong increase of the mean electron energy and number of ultra-relativistic electrons is reinforced by the results of the gamma-yield measurements that showed, in the case of pre-ionized polymer foams, 1000-fold increase of measured doses in all 10 channels of the gamma-spectrometer that covers the photon energy range from 30keV up to 100 MeV. For interpretation of the measured doses by means of the Monte Carlo multi-particle transport code FLUKA, the electron spectra were approximated by a Maxwell-like distribution with two temperatures. The best fit of the measured values obtained with a deviation better than 10% resulted in $T_1 \simeq$ 12 MeV and $T_2 \simeq$ 2-5 MeV, which is in a good agreement with direct measurements made by means of two electron spectrometers.

The experiment was supported by 3D-PIC simulations that accounted for the used laser parameters and the geometry of the experimental set-up and allow simulating the absolute number of accelerated electrons, their energy and angular distributions, which are in a good agreement with measurements. Simulations and measurements showed high directionality of the acceleration process, since the strongest increase of the electron energy, charge and corresponding gamma-yield was observed close to the direction of the laser pulse propagation.

The paper is organized as follows: laser and target parameters together with the used experimental set-up are described in Sec. I; angle dependent measurements of the electron spectral distribution and dosimetry of hard bremsstrahlung radiation are presented in Sec. II; in Sec. III PIC-simulations made for used experimental geometry are compared with experiment; Sec. IV summarizes experimental results and results of the modeling.

## 1. Experimental set-up

The goal of the performed experiments was to study the electron acceleration by intense sub-picosecond laser pulses propagating in sub-mm long plasmas of near critical electron density (NCD) and characterization of the energy distribution of super-ponderomotive electrons. In [30, 31] it was shown that interaction of relativistic laser pulses with foam targets of near critical electron density provides an efficient conversion of the laser energy into relativistic electron beams with energies far above those predicted by Wilks [9] $T_h = mc^2\left(\sqrt{a_L^2+1}-1\right)$ where $a_L$ is the normalized vector potential, $a_L^2 = 0.73 I_{L,18} \lambda_\mu^2$, $mc^2 = 511\,\text{keV}$, $I_{L,18}$ is the laser peak intensity in units of $10^{18}$ W/cm$^2$, and $\lambda_\mu$ is the laser wave length in μm. Distinctive features of these beams are their high, up to 100 MeV, energy accompanied by a large number of electrons (tens of nC) and strong directionality. This prediction was realized experimentally by the application of under-dense polymer foams while keeping the laser parameters fixed. Our experiments were performed at the Nd:glass PHELIX-laser facility at the Helmholtzzentrum GSI-Darmstadt, Germany. A laser pulse of 1.053μm fundamental wavelength, 80-100 J energy and 750±250 fs pulse duration was focused onto the target by means of a 150 cm 90° off-axis parabola under 5 degree to the target normal so that 60% of the laser energy was concentrated in the focal spot with a FWHM-size of 14±1×19±1 μm$^2$. Shot-to-shot deviations of the laser energy and 30% uncertainties in the laser pulse duration result in the rather large confidential interval of the laser intensities from 2.1 up to 5.1×10$^{19}$ W/cm$^2$ with corresponding normalized vector potentials $a_L$= 3.9-6.0. The ns laser contrast was kept at the highest level between 10$^{-11}$ and 10$^{-12}$.

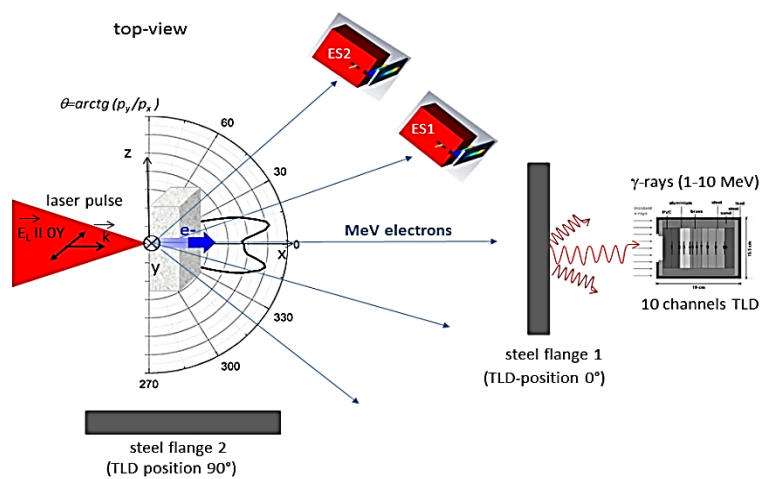

Figure 1. Experimental scheme: orientation of the laser beam onto the target; electron spectrometers ES1 and ES2 placed in the horizontal plane under 18° and 44° to the laser pulse propagation direction along the x-axis; and a ten channel TLD-detector for characterization of the gamma-radiation spectral distribution.

As targets low density polymer aerogels (foams) with 2 mg/cm$^3$ volume density and conventional metallic foils (Cu, Ti) were used. In order to create the NCD plasma, a foam layer/foil was irradiated by a 1.5 ns long pulse with a triangle temporal shape and 1-3 J energy forerunning the relativistic short main pulse. The delay between the peak of the ns pulse and the relativistic main pulse was varied from 0 up to 5ns.

The top-view of the experiment is presented in figure 1. The diagnostics set-up included two portable electron spectrometers with a 0.25 T static magnets. The spectrometers were placed in the target chamber at a distance of 450mm from the interaction point under 18° (ES1) and 44° (ES2) to the laser pulse propagation direction (x-axis, figure 1).

A TLD (thermo-luminescence dosimetry) - based ten channel system was used for the spectrometry of the hard bremsstrahlung caused by MeV electrons interacting with a 17mm thick Fe-flange that separated the evacuated target chamber from the enviroment. Ten TLD-cards were placed inside a shielding cylinder with a collimator window between absorbers made from lower to higher Z materials of different thickness [36]. The incident x-rays penetrate different absorbers and cause the corresponding TLD-signal (dose) in all ten cannels. The spectrometer was designed for an energy range from 30 keV up to 100 MeV. The TLD-detector was placed for one set of shots in the direction of the laser pulse propagation and for another one in perpendicular direction to measure an angular dependence of the MeV bremsstrahlung radiation produced by supra-thermal electrons.

In order to produce high aspect ratio homogeneous plasma with a slightly under-critical electron density, 2 mg/cm$^3$ cellulose triacetate (TAC, $C_{12}H_{16}O_8$) layers with thicknesses of 300 and 500 μm were used [32-34]. TAC-layers are optically transparent and characterized by highly uniform 3D network structure with 1-2 μm pore size, 0.1 μm thick and 1 μm long fibers with density of approximately 0.1 g/cm$^3$, and small 0.5% density fluctuations on the focal-spot size area of 100x100μm$^2$. Due to their open cell structure, air contained by pores can be evacuated. The mean volume density of 2 mg/cm$^3$ TAC-foam corresponds to $1.7\times10^{20}$ atoms/cm$^3$ and the mean ion charge $Z_{mean}$= 4.2. In the case of all CHO-atoms being fully ionized it results in a maximum of the electron density of $7\times10^{20}$ cm$^{-3}$, which is slightly below the critical density ($10^{21}$ cm$^{-3}$) for the fundamental wave-length of the Nd:glass laser ($\lambda$=1.053 μm).

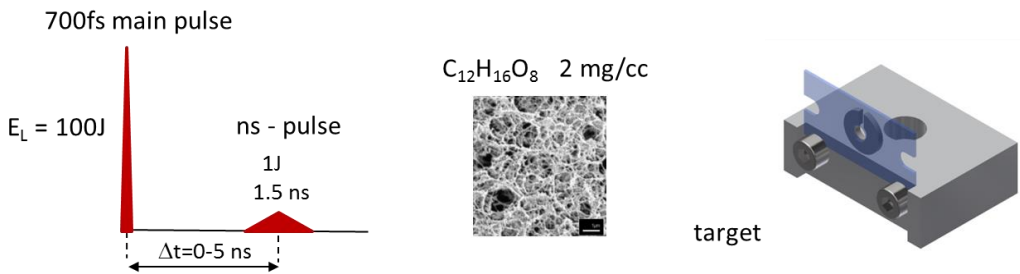

Figure 2. Combination of a low energy ns-pulse and sub-ps main-pulse; high resolution image of the TAC-structure; and a target station used in the experiment.

When the laser interacts with the foam surface, it starts to heat and ionize the solid membranes/fibers. The plasma created due to ionization of 0.1 g/cm$^3$ ($8\times10^{21}$ atoms/cm$^3$) dense 100nm thick fibers has an overcritical electron density and the fiber thickness is larger than the thickness of the skin layer (~ 30nm). Therefore it takes time until the created plasma will expand into the pore, reach under critical electron density and allow for further propagation of the laser pulse into the 3D aerogel-like structure. Intensity of the ns-pulse can be matched to the target density and target thickness in such way that the velocity of the ionization front will be much faster than the ion acoustic velocity and during the propagation of the supersonic ionization wave the heated high aspect ratio plasma region does not undergo notable expansion.

Analytically [37] and experimentally [38] it was shown, that the velocity of the super-sonic ionization front propagation in an initially porous matter is smaller than in an homogeneous medium of sub-critical density, like gas. Estimations of the ionization front propagation velocity $V_p$ that account for delay caused by hydro-homogenization processes on the micrometer scale have been made in accordance with eq. (27) [37]

$$V_p \approx 9.7 \times 10^4 \frac{A^{2/3} I_{14}^{1/3}}{Z_m^{2/3} \lambda_L^{4/3} \rho_a^\alpha \rho_s^{1-\alpha}}$$

for the following values that describe the laser parameters and $C_{12}H_{16}O_8$ foam structure used in the experiment: mean $A = 8$; mean ion charge $Z_m = 4.2$, mean foam density $\rho_a = 2$ mg/cm$^3$, fiber density $\rho_s = 0.1$ g/cm$^3$, laser wave length $\lambda_L = 1$ μm, normalized to $10^{14}$ W/cm$^2$ laser intensity $I_{14}=0.5$, and a power factor $\alpha = 0.8$ that reflects the geometry of the foam structure [37]. The velocity of the ionization front amounts $V_p = 2 \times 10^7$ cm/s and an estimated time required for ionization of a 300-500 μm thick foam layer by the 1 J ns-pulse reaches 1.5-2.5 ns. Due to the fact that the foam-targets were enclosed into Cu-washers (see target station in figure 2) we couldn't measure experimentally the electron density profile to the time of relativistic pulse arrival. We can assume that at large delays between the ns- and main pulses, the front part of the created plasma expands toward the incident laser. PIC-simulations made for a plasma layer with a constant NCD electron density and for a combination of the linear density profile with a density plateau showed no noticeable difference in the electron energy distribution and charge.

In the case of metallic foils, an inhomogeneous plasma plume that expands toward the main laser pulse with an exponentially decaying electron density was produced by the same 1-3 J ns pulse forerunning the relativistic short main pulse with delay varied from 0 up to 5 ns.

## 2. Experimental results

The laser shots that we analyse here were made onto Cu-foils and CHO-foams at $10^{-11}$ ns laser contrast with and without additional nanosecond pulse. Results on electron spectra measured by two electron spectrometers ES1 and ES2 and the TLD-dose caused by the bremsstrahlung radiation were very stable from shot to shot for every type of laser pulse - target combination.

*2.1 Electron spectra*

For evaluation of the electron energy distribution measured by two spectrometers with 0.25T static magnetic field we used a simulated dispersion curve. The principal scheme of the electron spectrometer is shown in figure 3. The spectrometer consisted of an iron housing that held two imaging plates (IPs). A signal on the long, side-on IP corresponds to electron energies 1< E <12 MeV and the short, end-on IP covers the electron energy range from 12 MeV up to 100 MeV. The response of BASF MS IPs to electron impact in dependence on the electron energy and angle of incidence was taken from [39, 40]. Raw electron spectra for high contrast shots onto Cu-foil (sh.30), shot onto the Cu-foil with a well-defined ns-pulse (sh. 27), and onto 300μm and 500 μm thick CHO-foam targets (sh. 31, 34, 38, 44) pre-ionized by the ns-pulse are presented in figure 4. Signals measured by the electron spectrometer ES1 placed 18° to the laser pulse propagation direction are shown on the right side of figure 4 and those measured by ES2 under 44° on the left side.

Two spots in the middle of the end-on IP (see e.g. figure 4, sh.31, ES1) are produced by x-rays (low intensity signal) and MeV-protons (saturated signal) that experience only a slight deviation in the 0.25 T magnetic field. The position of the x-ray signal would correspond to infinitely large electron energy.

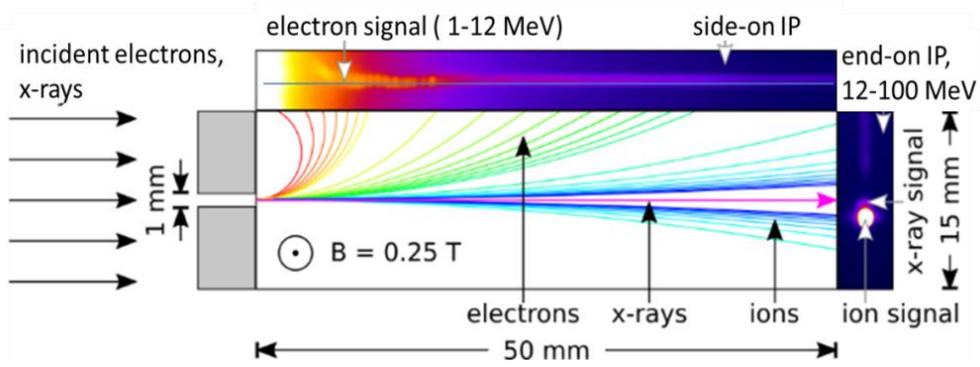

Figure 3. Scheme of the electron spectrometer.

In figure 4, one observes low background signal on the long side-on IP (electron energies $E \lesssim 12$ MeV) for shots onto foils at high laser contrast (sh. 30) and increased amplitude of the IP signal and higher electron energies in the case of a shot onto pre-ionized Cu-foil (sh.27).

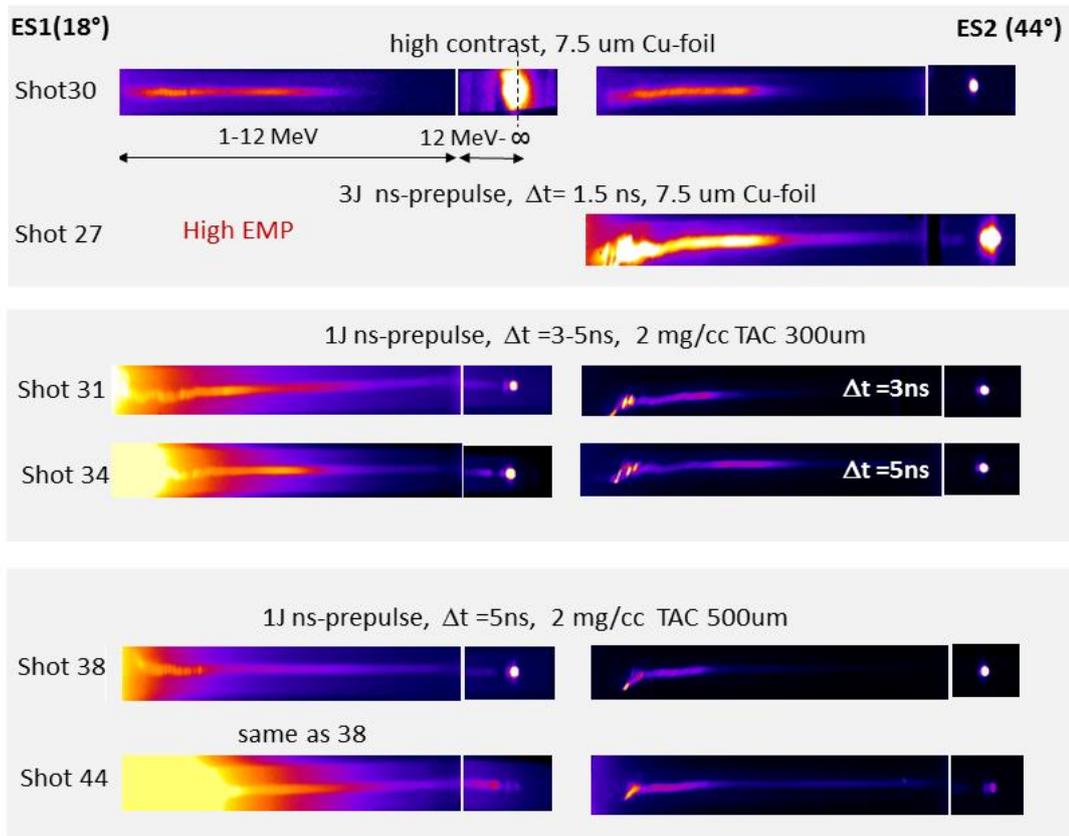

Figure 4. Raw electron spectra for various combinations of the ns-pulse energy, the delay between the ns- and short pulses $\Delta t$ and target parameters measured by means of a 0.25T static magnetic field and image plates (IPs) as detectors. Electron spectra were measured by means of electron spectrometers ES1 and ES2 under 18° and 44° to the laser propagation direction.

The saturated IP signal in the case of the laser irradiation of foams (sh. 31, 34, 38, 44), which was regularly detected in the IP-region with electron energies below 1 MeV, is probably caused by a strongly increased amount of laser accelerated electrons with $E > 6$-$8$ MeV. These electrons with a large stopping range penetrated directly the 5mm thick front iron-wall of the electron spectrometer and were deflected by magnets in accordance with their final energy. We excluded this energy region from our analyses and present results for electrons that entered the spectrometer through the 1mm collimator hole with energies above 4 MeV. A rather poor dispersion of the end-on IP, where signals from electrons with energies between 12 and 100 MeV are accumulated, still allowed for registration of the pronounced effect of the effective laser energy coupling into electrons in NCD-plasmas and their acceleration up to tens of MeVs (see raw electron spectra measured by ES1 end-on IP for shots 31, 34, 38, 44).

Data on absolute calibration of the image plates for relativistic electrons with energies up to 10 MeV and the dependence of the absolute signal on the electron angle of incidence onto the IP [39, 40] were used to evaluate the number of electrons in different energy ranges. Additionally, the IP-signal was corrected for energies above 10 MeV where the stopping power increases by a factor 2 between 10 and 100 MeV due to radiation losses. Experimental energy distributions of electrons with energies between 4 and 80 MeV that entered the spectrometer through the 1mm collimator hole are presented in figure 5(a). Figure 5(b) shows a total number of electrons that entered the spectrometer with energies above given one.

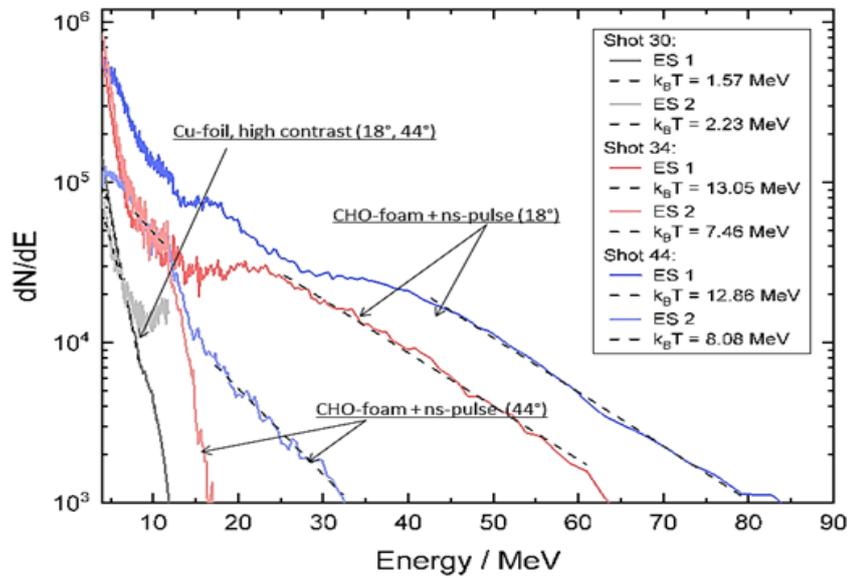

(a)

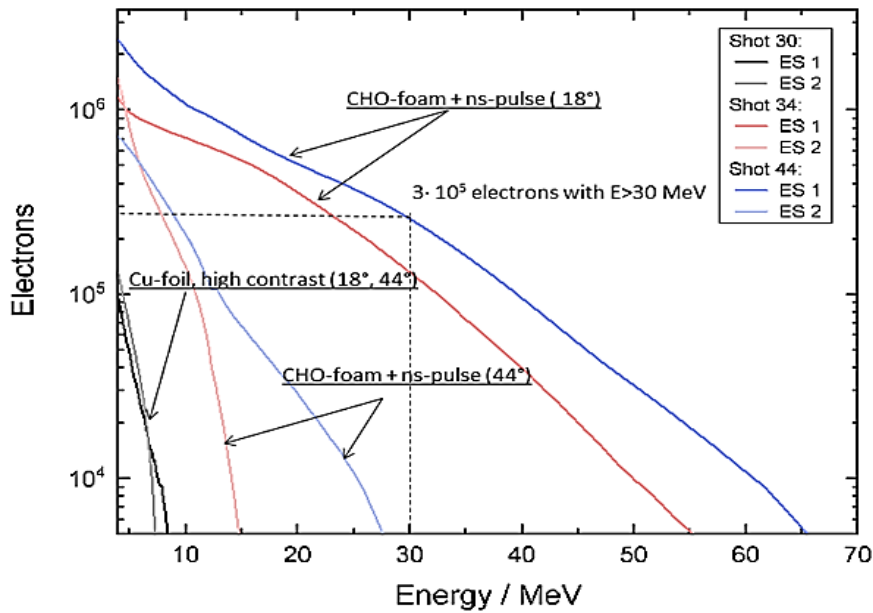

(b)

Figure 5. (a) Energy distribution of electrons with energies from 4 up to 80 MeV that entered the spectrometer; (b) Evaluated total number of electrons that enter the spectrometer with energies above given one.

Measured electron spectra were fitted by several Maxwell-like distributions with different temperatures (dashed lines in figure 5(a)). Part of the electron spectra that can be well approximated by a ponderomotive "temperature" of 1.5-3 MeV is present in all shots onto the foil and CHO-foams under 18° and 44°. When shooting onto the pre-ionized foams, we observe at least 10-fold increase of the number of hot electrons with $E < 10$ MeV measured by ES1 and ES2 and a dramatic increase of the number of ultra-relativistic electrons with energies above 20 MeV (sh. 34, 44, figure 4) measured by means of ES1 under 18°.

The maximum $T_h \simeq 13$ MeV was achieved in the case of 300μm and 500μm thick pre-ionized CHO-foam layers of 2 mg/cm$^3$ mean density and 5 ns delay between the pre-pulse and the main pulse (e.g. sh. 34, 44). This is at least 5-8.6 times higher than predicted by Wilks scaling determined by the vacuum laser pulse intensity 2.1- 5.1 ×10$^{19}$ W/cm$^2$ ($T_h$ = 1.55 - 2.60 MeV) and measured in the high contrast laser interaction with Cu-foils ($T_h$ = 1.57-2.23 MeV, e.g. sh. 30). The energy distribution of electrons detected under 18° and 44° is very similar in the case of shots with a high laser contrast (compare raw spectra and electron energy distribution measured by ES1 and ES2 in sh.30, figure 4). In interaction of the relativistic laser pulse with low density foams pre-ionized by the ns-pulse, we observe a predominant effect in the electron spectra measured by ES1 (figures 4, 5) under 18° to the laser propagation direction. An approximation of the high energy tails ($E > 25$-30 MeV) with a Maxwell-like distribution function results into 12.8-13 MeV electron temperature, while the measured amount of electrons with energies E > 30 MeV that entered the spectrometers trough the 1mm hole reached $3\times10^5$ particles (figure 5b). On the electron spectra measured by ES2 under 44°, the IP signal caused by electrons with $E > 25$-30 MeV is at the background level (figures 5a, b).

*2.2 Measurements of the TLD-doses caused by Bremsstrahlung radiation*

The observed tendency towards the strong increase of the mean electron energy and number of MeV laser-accelerated electrons is reinforced by the results of the gamma-yield measurements. The bremsstrahlung radiation (BS) was produced by MeV electrons passing through the 17 mm-thick steel flange, placed 868 mm apart from the target in the laser pulse direction, and was measured by means of an absolute calibrated 10-channel TLD-spectrometer [36] based on the thermo-luminescence dosimetry. Ten TLD-cards were placed between absorbers of different materials of increasing Z from PVC to steel and different thicknesses inside of a shielding cylinder with a collimator window. The materials of the TLD cards are pieces of doped lithium fluoride in two variations TLD 700 (7LiF: Mg, Ti) and TLD 700H (7LiF: Mg, Cu, P). They are suitable for detection of fast pulsed radiation. The response of the detectors is based on the excitation of the decoupled atoms. TLDs absorb radiation and emit photons proportionally to the deposited dose when heated to a few hundred degrees Celsius and have reduced residual signal. TLD 700H cards were calibrated for low- to high-dose ranges from 1 μGy to 20 Gy and have a very good dose-response and linearity at much higher doses. The spectrometer is designed for an energy range from 30 keV to 100 MeV, resolved in 10 different energy bins. The widths of the ten energy bins are: ΔEj = [0.02; 0.05; 0.15; 0.25; 0.5; 1.5; 2.5; 5.0; 40; 50] MeV with the interval centers: Ej = [0.04; 0.075; 0.175; 0.375; 0.750; 1.75; 3.75; 7.5; 30.0; 75.0] MeV, here j=1-10.

Figures 6 (a, b) show the measured dose readings that were obtained by the 10-channel TLD-spectrometer for laser shots generated at different conditions (target type, high contrast interaction or application of the ns-pulse in front of the relativistic main pulse, position of the TLD-spectrometer: (a) at 0° and (b) at 90° to the laser pulse propagation direction. When the TLD-spectrometer was placed at 0°, the lowest dose values were measured for shots onto Cu-foils (sh. 4, 25, 28) and 500μm thick CHO-foams (sh. 37) that used the high ns contrast ≲10$^{-11}$, see figure 6 (a). For high contrast laser shots no dependence of the TLD-doses on the target material and the target structure was measured. Shots 1, 2 deal with the case when the main laser pulse interacts with expanded plasma created by the ns-pulse that hit the Cu-foil.

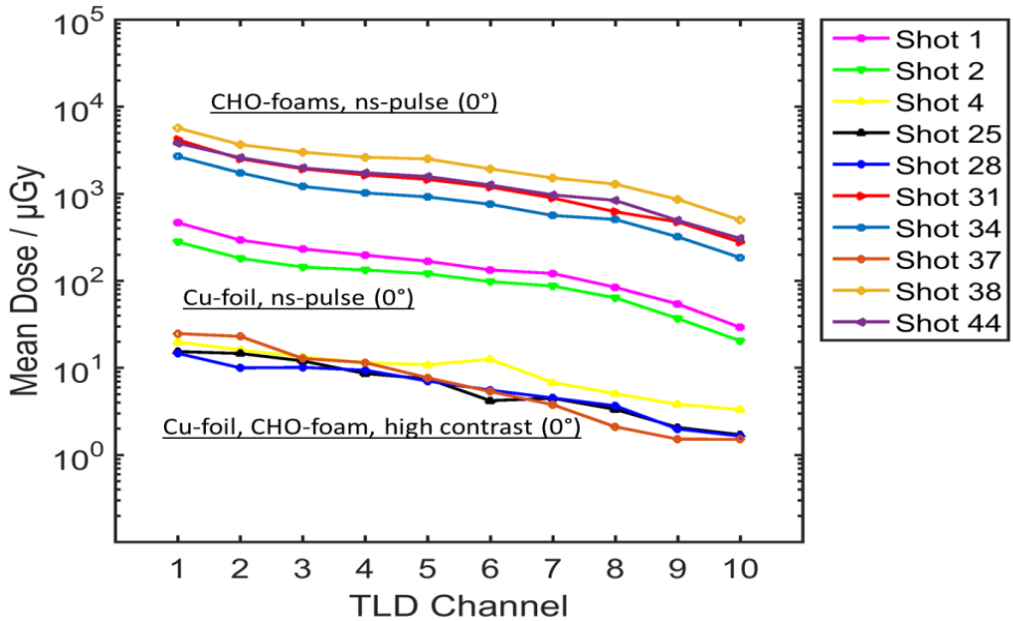

(a)

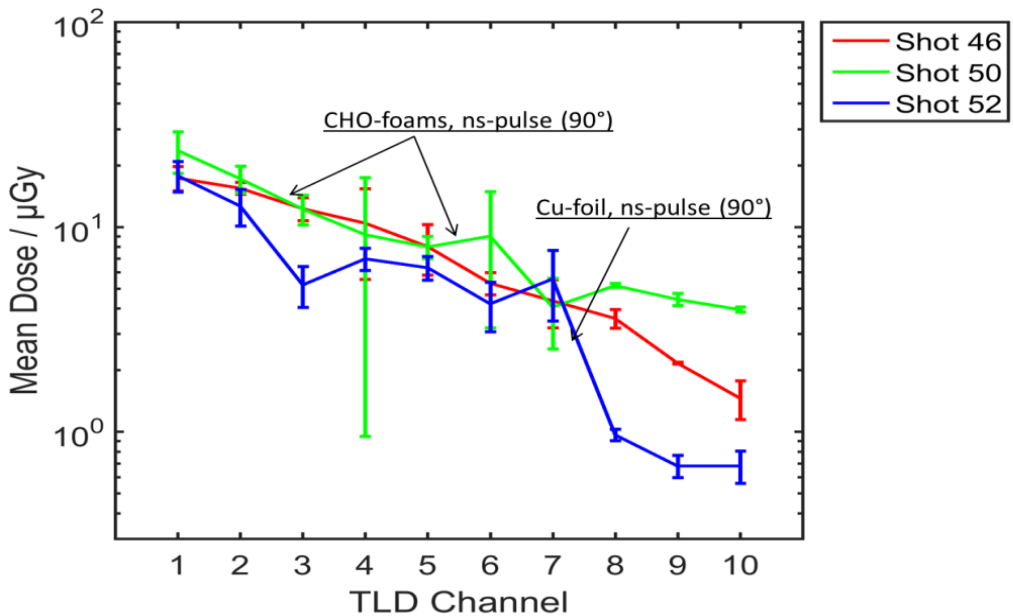

(b)

Figure 6. (a) Doses registered by the 10-channel TLD-spectrometer placed in the direction of the laser pulse propagation for shots made at different conditions; (b) same measurements made in the direction perpendicular to the laser pulse propagation.

The density profile of expanded plasma comprises a relatively short NCD-region where effective electron acceleration takes place. Experimentally we observe a strongly increased level of EMPs in the laser-bay outside the target chamber and one order of magnitude increase in the dose of the gamma-radiation measured by the TLD-detectors.

In relativistic laser interaction with pre-ionized CHO-foams, creation of the 300-500 μm long NCD plasmas ensures a longer acceleration path than in the case of freely expanding plasmas and results into even more effective coupling of the laser energy into energetic electrons. This effect was clearly observed e.g. in shots 31, 34, 37, 38, 44, where the relativistic laser pulse interacted with pre-ionized CHO-foam. TLD-measurements made in the direction of the laser pulse after propagation through the long scale NCD-plasma showed a 1000-fold increase of the measured doses in all 10 channels compared to the high contrast case.

In the perpendicular direction (figure 6 (b)), the TLD doses in the case of pre-ionized Cu-foil (sh. 52) and CHO-foam layers (sh. 46, 50) are measured to be very similar. The dose values are close to those measured in the direction of the laser pulse propagation at high laser contrast (figure 6 (a)).

The dose readings measured in the different channels of the TLD-spectrometer result from contributions made by photons and electrons with different energies. Consequently, the deconvolution of the spectral distributions of electrons requires information about the response functions of the dose of all the spectrometer layers to mono-energetic particle fluxes. The response matrix $R_{ij}$ was calculated using the Monte Carlo multi-particle transport code FLUKA [41, 42] in the energy region between 100 keV and 100 MeV for electrons. The real geometry of the experimental set-up and environment were recreated for the simulations with help of the FLAIR interface for FLUKA. The approximate dose values for 20 energy intervals with different interval widths $\Delta E$ = [0.1; 0.175; 0.25; 0.375; 0.5; 0.75; 1; 1.75; 2.5; 3.75; 5; 7.5; 10; 15; 20; 25; 30; 40; 50; 75; 100] MeV were calculated by the equation $D_{i\ calc} = \sum_{j=1}^{20} \Phi_j(E) R_{ij} \Delta E_j$ with an electron fluence $\Phi_j(E)$ and an average response $R_{ij}$ of a channel i over the energy interval $\Delta E_j$. For the calculation of the electron spectrum, an unfolding-algorithm was created and applied, since the analytical calculation is impossible due to an inverse problem. The electron fluence in dependence on energy was approximated by a Maxwell distribution function with two electron temperatures $T_1$ and $T_2$ and corresponding absolute number of electrons $N_1$ and $N_2$: $dN/dE_j = N_1/T_1 \times \exp(-E_j/T_1) + N_2/T_2 \times \exp(-E_j/T_2)$. The unfolding-algorithm was based on a sequential enumeration of matching data series and performed a best possible curve matching with the calculation of errors for deviations between experimental and simulated dose values. According to the above-written equation the spectral fluence was retrieved with a precision better than 10%.

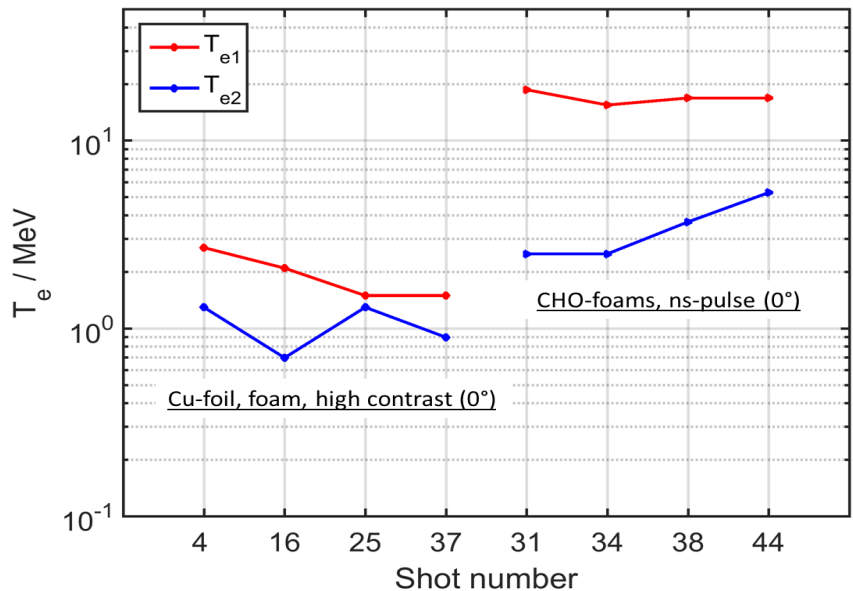

Figure 7. Electron temperatures evaluated from the measured TLD-doses via the best-fit procedure using FLUKA-simulations that account for the real geometry of the experimental set-up and environment and use two electron temperatures approximation. The best fit was achieved by keeping the deviations of the simulated doses from measured values below 10% for all 10 detector channels simultaneously.

Figure 7 shows the resulting values of $T_{e1}$ and $T_{e2}$ for selected laser shots: $N°$ 31, 34, 38, 44 were made onto pre-ionized foam and 4, 25, 28, 37 onto foil/foam at highest laser contrast. As expected, in the case of foams, when the TLD doses reached their maximum value, both electron temperatures are essentially higher compared to $T_{e1}$ and $T_{e2}$ evaluated for the case of high contrast shots. These results show a very good correlation with the experimental conditions and with the results of the electron spectrometers. Measured doses allow retrieving an absolute number of electrons $N_1$ and $N_2$ that interacts with the flange and produce a signal inside the TLD-cards. According to the set-up geometry, these are electrons that propagate along the x-axis with a divergence $\theta = 3.3°$ (half angle). In the case of the laser interaction with pre-ionized foams, the best fit of all ten TLD-signals was obtained for $T_1 \simeq 12$ MeV and $N_1 = 0.5$-$1 \times 10^{10}$ electrons. This amount corresponds to up to 8-16 nC of the well-directed super-ponderomotive electron beam.

## 3. Results of modelling and comparison with experimental data.

3D PIC simulations were performed using the Virtual Laser Plasma Laboratory code [43] for the laser parameters and interaction geometry used in the experiment (Section I, figure 1). In particular, the FWHM axes of the elliptical laser focal spot were 13.4 and 18.8 μm, the laser pulse energy in the focal spot was 54 J (full pulse energy 94 J) that at the FWHM pulse duration of 700 fs corresponded to the laser intensity of $4.4 \times 10^{19}$ W/cm$^2$ ($a_L = 5.67$). The plasma was composed of electrons, fully ionized ions of carbon, hydrogen and oxygen. Simulations accounted for the ion type and the ion fraction in accordance with the chemical composition of triacetate cellulose $C_{12}H_{16}O_8$, see e.g. [30]. The simulation box had a size of 610 μm along the x-axis. The first 10 and the last 100 μm of the space in this direction were free of the plasma at the initial moment. The box had 100 μm size both along the y-axis and the z-axis. Sizes of a numerical cell were 0.05 μm along the x-axis and 0.5 μm along the y-axis and the z-axis. The number of particles per cell equalled 4 for the electrons and 1 for the ions of each type. Boundary conditions were absorbing for particles and fields in each direction.

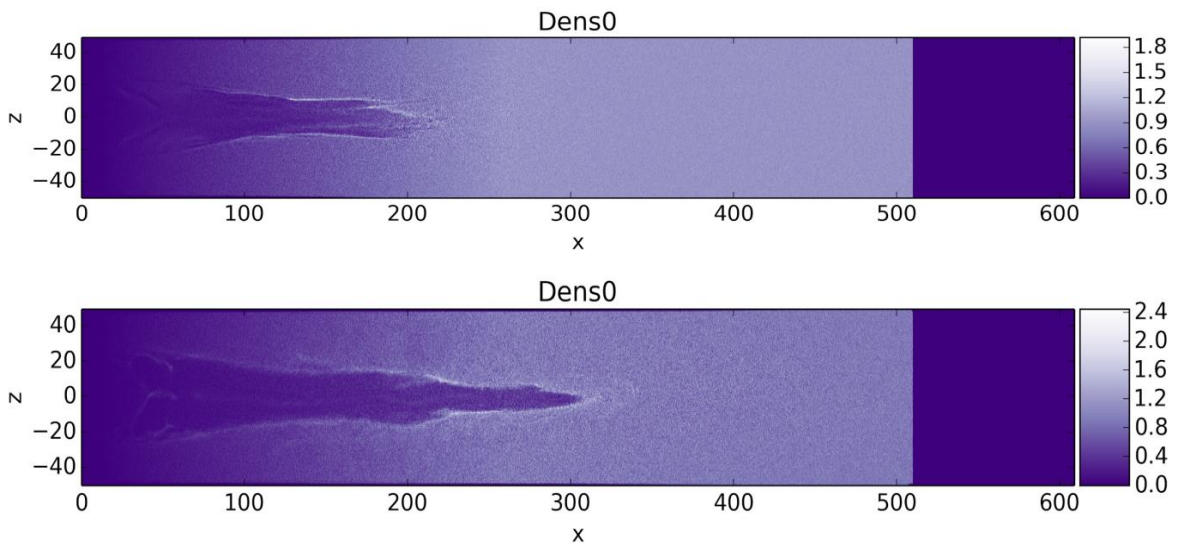

Figure 8. Snapshots of the electron density distributions in the plane XZ (see fig.1) at the time moments corresponded to ct = 150 and 550 μm (t = 0 corresponds to the laser pulse maximum at the target front side). The electron density is presented in $n_e/n_{e\text{-}cr}$.

In order to imitate the plasma expansion toward the laser before the main short intense laser pulse strikes the target, the initial electron density (together with the neutralizing ion density) was presented as a combination of the linear increasing density profile and a density plateau ($n_e = (0.1+0.9 (x-10)/250) n_{cr}$ for $x < 260$ μm and $n_e = n_{cr}$ for $x \geq 260$ μm). Figure 8 illustrates the electron plasma density dynamics during the laser pulse propagation in the plasma channel. Strong self-focusing, filamentation and bifurcation of the laser pulse, which are clearly seen in this figure, are reflected in the angular distributions of accelerated electrons shown in figure 9.

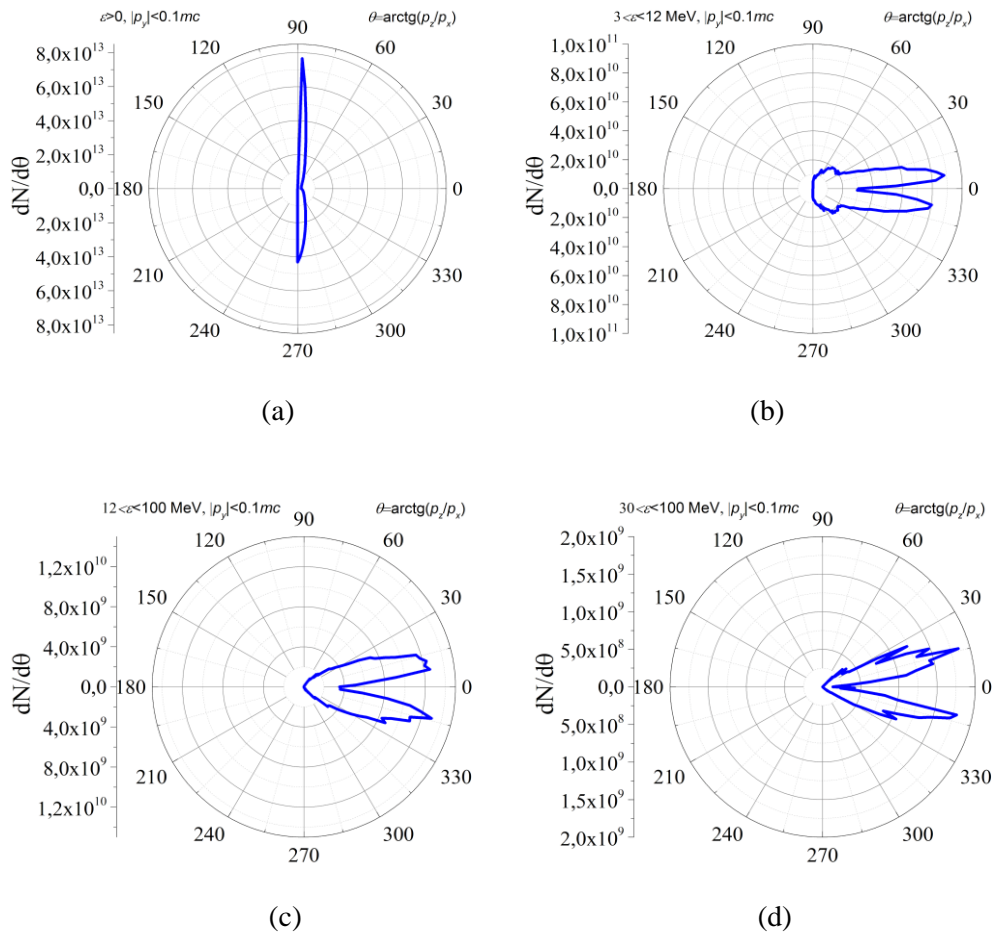

Figure 9. Distributions over angle $\theta = arctg(p_z/p_x)$ of electrons, which left the target with $p_x > 0$ and $|p_y| <$ 0.1$mc$, at the time moment corresponding to $ct$ = 550 μm for different energy ranges: (0, ∞) (a); (3, 12) MeV (b); (12, 100) MeV (c) and (30, 100) MeV (d).

A comparison of the angular distribution of accelerated electrons without restriction on the energy (depicted in figure 9(a)) with distributions at higher energies, figure 9(b), (c) and (d), shows that the main part of relatively low energy electrons (E < 3 MeV) are accelerated mostly in the radial direction by a ponderomotive force that pushes electrons out of the plasma channel. Super-ponderomotive electrons (E > 3 MeV) that experience the betatron oscillatons and the direct laser acceleration in the plasma channel [25] leave the plasma under the angle to the direction of the laser propagation axis *OX*; this angle is determined by the ratio of the transversal electron momentum $p_y$ and $p_z$ to the longitudinal $p_x$. Increasing of the angle under which electrons with higher energies move relative to the laser propagation axis *OX* (compare figure 9 (b) with (c) and (d)), reflects an increased transverse momentum of electrons accelerated by the laser pulse due to transverse betatron oscillations in the self-generated quasi-static electric and magnetic fields of the plasma channel. The presence of self-generated fields is the characteristic feature of a laser pulse channelling in plasma caused by the relativistic and ponderomotive effects. Ponderomotive expulsion of background plasma electrons from the channel creates a radial quasi-electrostatic field, while the current of the accelerated electrons generates an azimuthal magnetic field [25, 26, 30, and 31]. Addi-tional injection and acceleration of electrons that occurs also in the surface waves generated near the plasma channel walls was discussed in [22, 26] on the base of 2D PIC simulations for the case of subcritical plasmas. The presence of such waves in the near-critical plasmas was observed in full 3D PIC-simulations of the density and field distributions, see e.g. [30, 31].

We would like to emphasize that the 3D capability of the PIC code allows simulations that are close to real experimental conditions. Thus, the absolute energy spectra, i.e., the number of accelerated electrons in any energy range, and also its angular distribution can be obtained. Electron spectra were simulated at the ranges of angles $\theta = arctg(p_z/p_x) = 18°\pm5°$ (magenta line) and $44°\pm5°$ (blue line), which correspond to the positions of the ES1 and ES2 spectrometers (figure 10). For $4.4\times10^{19}$ W/cm$^2$ ($a_L$ = 5.67) vacuum laser intensity, high energy parts of the electron energy distributions (E > 25 MeV) simulated under 18° and 44° to the laser pulse propagation direction were approximated by the Maxwell-like functions with hot electron temperatures $T_h$ = 17.7 MeV and 12.2 MeV. Note that in our simulations we used slightly smaller pulse duration, 700 fs, than the average value of 750±250 fs indicated for the range operated in the experiment. Increasing of the pulse duration together with some decreasing of the pulse energy in the confidence intervals will lead to decrease of the laser intensity and consequently to lower hot electrons temperatures as it was checked in simulations. The squareroot scaling of the $T_h$ with the laser intensity suggested in [25] would lead to $T_h$ = 12 MeV ($\theta = 18°\pm5°$) and 8.6 MeV ($\theta = 44°\pm5°$) for $2.1\times10^{19}$ W/cm$^2$ that represents the low limit of the confidential interval of laser intensities used in the experiment (see Sec. 1). These numbers are in a good agreement with direct measurements of the electron energy distribution resulting in $T_h$ = 12.8-13 MeV ($\theta = 18°\pm5°$) and 7.5-8.0 MeV ($\theta = 44°\pm5°$), see figure 5a.

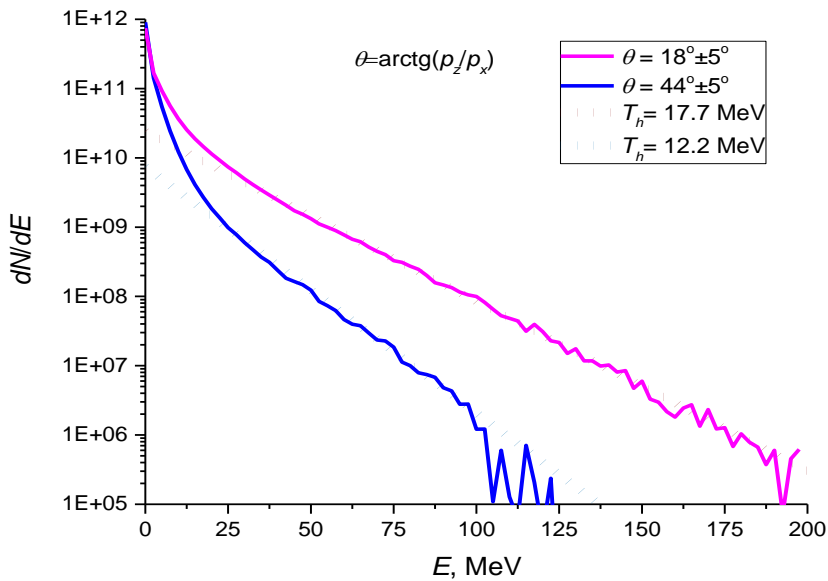

Figure 10. Electron energy spectra obtained in PIC simulations at $4.4\times10^{19}$ W/cm$^2$ ($a_L$ = 5.67) for spectrometers ES1 and ES2 at the ranges of angles $\theta = arctg(p_z/p_x) = 18°\pm5°$ (magenta line) and $44°\pm5°$ (blue line). Brown and green dotted lines are exponential fits for temperatures $T_h$ = 17.7 MeV and 12.2 MeV.

The 3D geometry of the simulations that was chosen in conformity with the experimental set-up allows obtaining not only the shape of the spectra and realistic angular distributions, but also the absolute numbers of accelerated particles. Taking into account the solid angle, under which the electron spectrometer ES1 is seen, the number of electrons reaching the spectrometer with energies above 30 MeV is 6 $\times10^5$, as follows from Fig. 9(d). Similar number of electrons can be estimated from figure 10 for the range of angles $\theta = arctg(p_z/p_x) = 18°\pm5°$ (magenta line), taking into account that the energy spectrum accumulates particles with any transverse momenta $p_y$. The obtained number of electrons with energies above 30 MeV, 6 ×10$^5$ coincides rather well with the number measured by the electron spectrometer ES1 in shot 44 (3 ×10$^5$, see figure 5(b)). Table 1 presents the total number of electrons, their charge and corresponding conversion efficiency for different intervals of electron energies obtained in PIC-simulations. The total charge of electrons with energies above 3 MeV reaches 2 µC, while the charge of super-ponderomotive electrons with E > 30 MeV reaches a very high value of 78nC.

| Energy range, MeV | Number of electrons | Charge of electrons, µC | Percent of laser energy |
|---|---|---|---|
| (0.5, +∞) | 3.22×10$^{13}$ | 5.15 | 40% |
| (3, +∞) | 1.15×10$^{13}$ | 1.84 | 31% |
| (3, 12) | 9.09×10$^{12}$ | 1.45 | 14.8% |
| (12, +∞) | 2.38×10$^{12}$ | 0.381 | 16.3% |
| (30, +∞) | 4.85×10$^{11}$ | 0.078 | 6.4% |

Table 1. Energy dependent number, charge and conversion efficiency of accelerated electrons which left the target with $p_x$ > 0 at ct=550 µm predicted by PIC-simulations.

### 4. Conclusion

In this article we present new experimental results on the interaction of relativistic sub-picosecond laser pulses with extended, sub-mm long NCD-plasmas. Low density polymer foam layers were used to create this type of hydrodynamic stable, large scale, quasi-homogeneous plasmas. Interaction of the relativistic laser pulse with large-scale NCD-plasmas ensures a long acceleration path and results into effective coupling of the laser energy into energetic electrons.

Experiments on the electron heating by a 80-100J, 750 fs short laser pulse of 2-5×10$^{19}$ W/cm$^2$ intensity demonstrated that the effective temperature of supra-thermal electrons increased from 1.5-2 MeV, in the case of the relativistic laser interaction with a metallic foil at high laser contrast, up to 13 MeV for the laser shots onto the pre-ionized 300-500 µm long foam layer with a near critical electron density. Measurements showed high directionality of the acceleration process.

The observed tendency towards the strong increase of the mean electron energy and number of super-ponderomotive electrons is reinforced by the results of the gamma-yield measurements. In the case of laser interaction with long-scale NCD-plasmas, the dose caused by the gamma-radiation measured in the direction of the laser pulse propagation showed a 1000-fold increase compared to the high contrast shots onto plane foils and doses measured perpendicular to the laser propagation direction for all used combinations of targets and laser parameters. The super-ponderomotive electron temperature retrieved from the measured TLD-doses by means of the Monte-Carlo simulations reached 11-12 MeV, which is in a good agreement with direct measurements by the static magnets.

The experiment was supported by the full 3D-PIC simulations that account for the used laser parameters and the geometry of the experimental set-up and allow, in contrast to usually used 2D-3V PIC analysis, simulating the absolute number of accelerated electrons, their energy and angular distributions. Obtained simulation results are in a good agreement with measured amount of electrons that were registered by the electron spectrometers and indicate the effective electron acceleration in NCD plasmas with a total electron charge of about 2 µC at the energies above ponderomotive one (> 3 MeV), while the charge of super-ponderomotive electrons with E > 30 MeV reaches a high value of 80 nC.

**Acknowledgments**

The experimental group is very thankful for the support provided by the PHELIX-team and the RFBR 17-02-00366 project.

**References**

[1] Pape S Le, Neumayer P, Fortmann C *et al* 2010 *Phys. Plasmas* **17** 056309

[2] Fedeli M L, Batani D *et al* 2014 *Phys. Plasmas* **21** 102712

[3] Tahir N A, Deutsch C, Fortov V E *et al* 2005 *Phys. Rev. Lett*. **95** 035001

[4] Tahir N A, Stöhlker Th, Shutov A *et al* 2010 *New J. Phys.* **12** 073022

[5] Ravasio A, Koenig M, Pape S Le *et al* 2008 *Phys. Plasmas* **15** 060701

[6] Li K, Borm B, Hug F *et al* 2014 *Laser Part. Beams* **32**(04) 631

[7] Brunel F 1987 *Phys. Rev. Lett.* **59** 52

[8] Wilks S C, Kruer W L, Tabak M and Langdon A B 1992 *Phys. Rev. Lett.* **69** 1383

[9] Gibbon P 2005, *Short Pulse Laser Interactions with Matter: An Introduction*. Imperial College Press/World Scientific London/Singapore. ISBN 1-86094-135-4.

[10] Mulser P and Bauer D 2010 *High-Power Laser-Matter Interaction* (Springer, Heidelberg).

[11] Mulser P, Weng S M and Liseykina T 2012 *Phys. Plasmas* **19** 043301

[12] Bochkarev S G, Brantov A V, Bychenkov V Y *et al* 2014 *Plasma Phys. Rep*. **40**, 202–214

[13] Andreev N E, Pugachev L P, Povarnitsyn M E, and Levashov P R 2016 *Laser and Part. Beams,* **34**, 115–122.

[14] Andreev N E, Kuznetsov S V, Cros B *et al* 2011 *Plasma Phys. Control. Fusion* **53** 014001

[15] Leemans W P, Gonsalves A J, Mao H S *et al* 2014 *Phys. Rev. Lett.* **113** 245002

[16] Esarey E, Schroeder C B and Leemans W P 2009 *Rev. of Modern Physics* **81** 1229

[17] Faure J, Glinec Y, Pukhov A *et al* 2004 *Nature* **431** 541–4

[18] Pugacheva D V, Andreev N E 2018 *Quantum Electronics* **48**(4) 291

[19] Walker P A, Bourgeois N, Rittershofer W *et al* 2013 *New J. Phys.* **15** 045024-17

[20] Ju J, Genoud G, Ferrari H E *et al* 2014 *Phys. Rev. ST Accel. Beams* **17** 051302

[21] Willingale L, Nilson P M, Thomas A G R *et al* 2011 *Phys. Plasmas* **18** 056706

[22] Willingale L, Thomas A G R, Nilson P M *et al* 2013 *New J. Phys.* **15** 025023

[23] Toncian T, Wang C, McCary E *et al* 2016 *Matter and Radiation at Extremes* **1** 82-87

[24] Willingale L , Arefiev A V, Williams G J *et al* 2018 *New J. Phys*. **20** 093024

[25] Pukhov A, Sheng Z-M and Meyer-ter-Vehn J 1999, *Phys. Plasmas* **6** (7) 2847

[26] Arefiev A V, Khudik V N, Robinson A P L *et al* 2016 *Phys. Plasmas* **23**, 056704

[27] Khudik V, Arefiev A, Zhang Xi *et al* 2016 *Phys. Plasmas* **23**, 103108

[28] Borghesi M, Mackinnon A J, Gaillaed R *et al* 1998 *Phys. Rev. Lett.* **80** 5137

[29] Gray R J, Carrol D C, Yaun X H *et al* 2014 *New J. Phys.* **16** 113075

[30] Pugachev L P, Andreev N E, Levashov P R, Rosmej O N 2016 *Nucl. Instr. Methods. A* **829** 88

[31] Pugachev L P and Andreev N E 2018 *J. of Phys. Conference Series* in press

[32] Khalenkov A M, Borisenko N G, Kondrashov V N *et al* 2006 *Laser Part. Beams* **24** (02) 283

[33] Borisenko N G, Akimova I V, Gromov A I *et al* 2006 *Fusion Science and Tech.* **49**(4) 676

[34] Borisenko N G, Khalenkov A M, Kmetik V *et al* 2007 *Fusion Science and Tech*. **51**(4) 655

[35] Doria D, Kar S, Ahmed H *et al* 2015 *Rev. Sci. Instrum.* **86** 123302

[36] Horst F, Fehrenbacher G, Radon T *et al* 2015 *Nucl. Instr. Methods. A* **782** 69

[37] Gus'kov Yu, Limpouch J, Nicolai Ph and Tikhonchuk V T 2011 *Phys. Plasmas* **18** 103114

[38] Nicolaï Ph, Olazabal-Loumé M, Fujioka S *et al* 2012 *Phys. Plasmas* **19** 113105

[39] Bonnet T, Comet M, Denis-Petit D *et al* 2013 *Rev. Sci. Instr.* **84** 103510

[40] Tanaka K A, Yabuuchi T, Sato T *et al* 2005 *Rev. Sci. Instr.* **76** 013507

[41] Böhlen T T, Cerutti F, Chin M P W *et al* 2014 *Nuclear Data Sheets* **120** 211-214

[42] Ferrari A, Sala P R, Fasso A, and Ranft J 2005 *CERN-2005-10, INFN/TC_05/11, SLAC-R-773*

[43] Pukhov A 1999 *J. Plasma Phys.* **61** 425–433